\documentclass[journal]{IEEEtran}
\usepackage{blindtext}
\usepackage{graphicx}
\usepackage{multirow}
\usepackage{booktabs}
\usepackage{comment}
\usepackage{xcolor}
\usepackage{amssymb}
\usepackage{mathtools}
\usepackage{hyperref}
\usepackage{tikz}
\usepackage{pgfplots}
\usepackage{color, colortbl}
\usepackage{amsmath}
\usepackage{url}
\usepackage{bm}
\usepackage{float}

\definecolor{LightCyan}{rgb}{0.95,1,1}

\definecolor{blue}{RGB}{0,0,255}

\begin{document}

\title{U-Netmer: U-Net meets Transformer for medical image segmentation}

\author{Sheng He, Rina Bao, P. Ellen Grant, Yangming Ou
       
\thanks{

%Manuscript received by XXXX, accted XXXX. 
(\textit{Sheng He and Yangming Ou are the corresponding authors.}) \\
S. He, R. Bao, P. Grant and Y. Ou are with the Boston Children's Hospital and Harvard Medical School, Harvard University, 300 Longwood Ave., Boston, MA, USA.
E-mail: heshengxgd@gmail.com; rina.bao@childrens.harvard.edu,
ellen.grant@childrens.harvard.edu, yangming.ou@childrens.harvard.edu}% <-this % stops a space
}

\maketitle

\begin{abstract}
The combination of the U-Net based deep learning models and Transformer is a new trend for medical image segmentation.
U-Net can extract the detailed local semantic and texture information and Transformer can learn the long-rang dependencies among pixels in the input image.
However, directly adapting the Transformer for segmentation has  ``token-flatten" problem (flattens the local patches into 1D tokens which losses the interaction among pixels within local patches) and ``scale-sensitivity" problem (uses a fixed scale to split the input image into local patches).
Compared to directly combining U-Net and Transformer, we propose a new global-local fashion combination of U-Net and Transformer, named U-Netmer, to solve the two problems. The proposed U-Netmer splits an input image into local patches. The global-context information among local patches is learnt by the self-attention mechanism in Transformer and U-Net segments each local patch instead of flattening into tokens to solve the `token-flatten" problem.
The U-Netmer can segment the input image with different patch sizes with the identical structure and the same parameter.
Thus, the U-Netmer can be trained with different patch sizes to solve the ``scale-sensitivity" problem.
We conduct extensive experiments in 7 public datasets on 7 organs (brain, heart, breast, lung, polyp, pancreas and prostate) and 4 imaging modalities (MRI, CT, ultrasound, and endoscopy) to show that the proposed U-Netmer can be generally applied to improve accuracy of medical image segmentation.
These experimental results show that U-Netmer provides state-of-the-art performance compared to baselines and other models.
In addition, the discrepancy among the outputs of U-Netmer with different scales is linearly correlated to the segmentation accuracy which can be considered as a confidence score to rank test images by difficulty without ground-truth.
The code will be available on GitHub.
\end{abstract}


\IEEEpeerreviewmaketitle

\section{Introduction}
\begin{figure}[!t]
\centering
\includegraphics[width=0.5\textwidth]{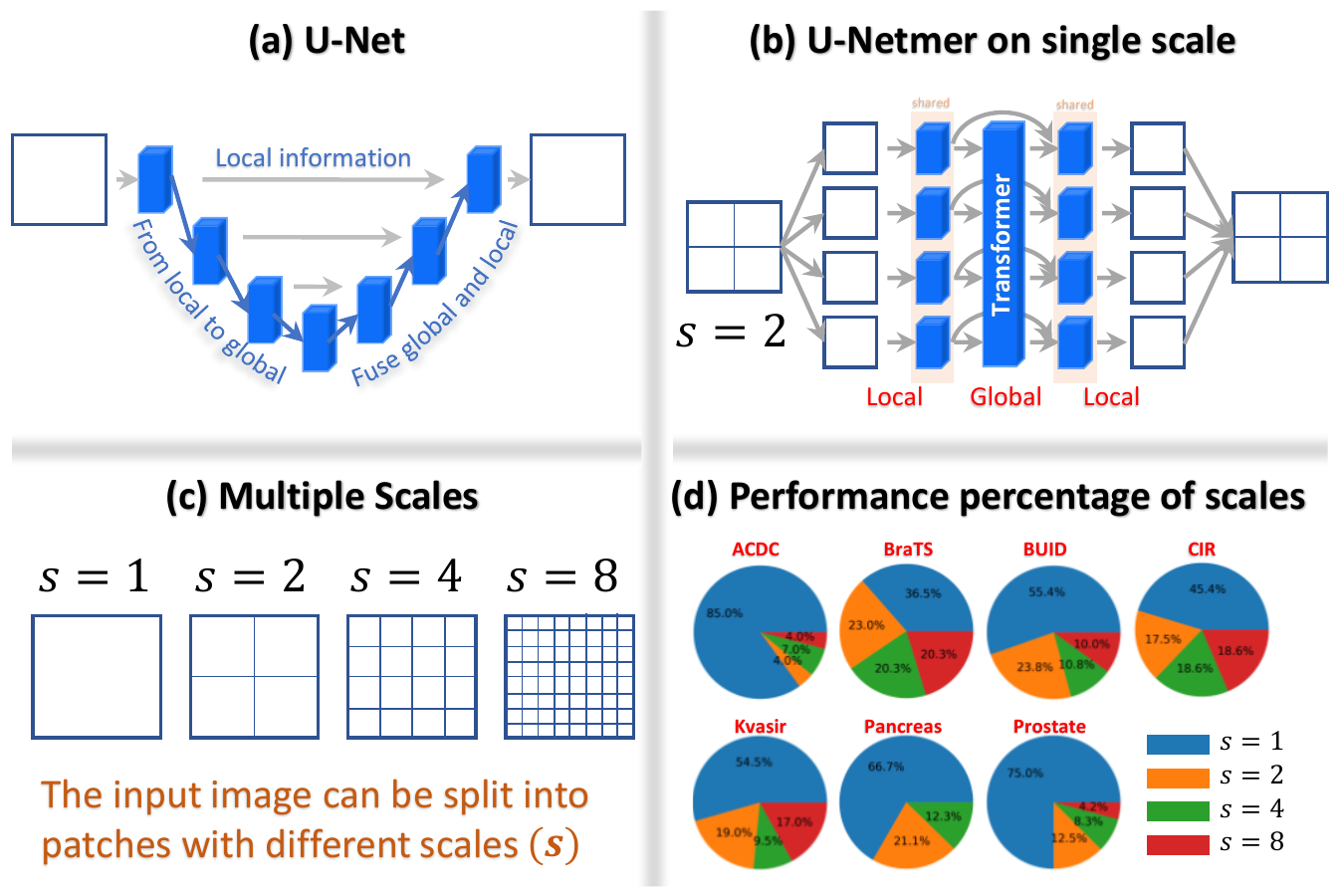}
\caption{Sketch of the proposed algorithm. (a) U-Net based models implicitly integrate local information by down-sampling the features. (b) The proposed U-Netmer model on single scale ($s=2$) explicitly splits the input image into 4 equal-size local patches and uses U-Net to segment each local patch and uses Transformer to fuse the local information among local patches. (c) The input images can be split into patches with different scales $s=1,2,4,8$. When $s=1$, the input is the whole image and $s=2,4,8$ means splitting the input image into $s^2$ equal-size and no-overlap patches. (d) The percentage of the best performance achieved with different scales $s$ on test samples of 7 public datasets.}
\label{fig:concepts}
\end{figure}

Medical image segmentation aims to use machine learning models (e.g., Convolutional Neural Networks or CNNs for short) to automatically segment the target regions (organs or lesions) from the input medical images with different modalities~\cite{zhou2019unet,schlemper2019attention,antonelli2022medical,isensee2021nnu}.
One popular backbone of the deep learning model for segmentation is U-Net~\cite{ronneberger2015u}, which is a general CNN model with an encoder and decoder structure (Fig.~\ref{fig:concepts}(a)).
The encoder path decomposes the input image from local to global deep features where the spatial size is gradually reduced by using the max-pooling operation. 
As the layer goes to deep, it extracts the high-level contextual information by discarding the detailed information on each local pixel to remove noise and irrelevant information~\cite{he2023segmentation}.
To recover the lost detailed spatial information, the decoder path hierarchically fuses global features from the output of the encoder and local features from the intermediate output of the encoder~\cite{ronneberger2015u} for computing the final segmentation map.
In summary, as shown in Fig.~\ref{fig:concepts}(a), the U-Net model implicitly extracts features from local information towards global contextual information and fuses these features gradually for the final segmentation output which has the same spatial size as the input image.

Based on the basic structure of the U-Net, many variations have been proposed~\cite{isensee2021nnu,azad2022medical}.
For example, U-Net++~\cite{zhou2019unet} uses multiple decoder paths on different scales to fuse the global and local information.
Inspired by the attention mechanism which can perform feature recalibration in deep neural networks~\cite{hu2018squeeze}, an attention module has been introduced in U-Net for medical image segmentation~\cite{schlemper2019attention}.
Inspired by the success of vision transformer~\cite{dosovitskiy2020image}, 
there is a new trend to integrate Transformer into U-Net to fuse information from different scales or resources to boost the performance of the U-Net~\cite{chen2021transunet,wang2022uctransnet,valanarasu2021medical}.
For example, MedT~\cite{valanarasu2021medical} uses a two-branch structure including a global branch to learn global information by CNN-based neural networks and a local branch to learn local information using Transformer.
Trans U-Net~\cite{chen2021transunet} applies the Transformer on the last layer of the encoder from the U-Net to boost the encoder of the U-Net.
UCTransNet~\cite{wang2022uctransnet} uses the Transformer to fuse the intermediate features of U-Net in different scales obtained after max-pooling operations.

Most models of combing the U-Net and Transformer follow the same structure of U-Net and consider the Transformer as a sub-module which can learn the long-range dependencies on deep features with different sizes to boost the performance of segmentation~\cite{azad2022medical}.
Few studies use the strategy of cutting the input image into local patches (which is also known as ``patchification"~\cite{beyer2022flexivit}) for medical image segmentation (as shown in Fig.~\ref{fig:concepts}(c)) which 
which raises at least two issues existed~\cite{valanarasu2021medical}:
(1) \textbf{``token-flatten issue"}: the vision Transformer flattens the local patches into 1D tokens, but the 1D functionality losses the interaction of the tokenized information on the local patches~\cite{wang2022uctransnet} and (2) \textbf{``scale-sensitivity issue"}: the vision Transformer usually uses a fixed scale to split the input image into patches and the performance of medical image segmentation is sensitive to the scale $s$ when cutting the input image into patches with different sizes (as shown in Fig.~\ref{fig:concepts}(c)). 
Fig.~\ref{fig:concepts}(d) shows the percentage of different scales which achieve the best performance on test images of 7 different datasets (the description of datasets can be see in Section~\ref{sec:datasets} and the experiment is described in Section~\ref{sec:pertange}).
It shows that not all test samples have the highest segmentation accuracy with scale $s=1$ and some test images achieve the best accuracy with other scales $s=2,3,4$.
For example, on BraTS, 36.5\% of the test images have the best segmentation accuracy with scale $s=1$, and 23.0\%, 20.3\%, 20.3\% of the test images achieve the best accuracy with scale $s=2,3,4$, respectively.
Thus, directly cutting the input image into patches with a fixed scale and feeding them into Transformer is not necessarily optimal for segmentation. 

In this paper, we propose a simple and efficient neural network to optimally combine the U-Net and Transformer for segmentation, named U-Netmer.
Similar to vision Transformer, it explicitly splits the input image into local patches and uses Transformer to integrate local information among these patches (shown in Fig.~\ref{fig:concepts}(b)).
The ``patchification" can provide many potential applications for segmentation inspired by the successful application for classification, such as information fusion from different modalities~\cite{diao2022metaformer}, patch dropping and reconstruction for self-supervised learning~\cite{he2022masked} and few-shot learning for dense prediction~\cite{Universal2023icrl}.

To solve the \textbf{``token-flatten issue"}, 
U-Netmer uses a backbone of a standard segmentation neural network (such as U-Net) to perform the segmentation on local patches instead of flatten each patch into 1D tokens.
A standard neural network usually contains \textit{an encoder} and \textit{a decoder} (as shown in Fig.~\ref{fig:concepts}(a)). 
The output deep feature of the encoder is reshaped into 1D tokens and all tokens from the local patches segmented from the input image are concatenated as a sequence of tokens as the input of a standard Transformer~\cite{vaswani2017attention}  (as shown in Fig.~\ref{fig:concepts}(b)).
The Transformer uses a self-attention mechanism to learn the global-contextual information among local patches to enhance the segmentation for each local patch.

To solve the \textbf{``scale-sensitivity issue"},
an identical U-Netmer with the same parameter is trained on local patches segmented with different scales $s=1,2,4,8$. 
Thanks to the flexible structure of the U-Netmer, it can be used on arbitrary patch sizes without any changes of the network structure (as shown in Fig.~\ref{fig:multiscale}).
Multi-scale patches are designed to reduce segmentation's sensitivity to patches at single scale~\cite{beyer2022flexivit}.

The main contributions of the work are summarized below:
\begin{itemize}
    \item We propose U-Netmer which consists of a backbone to extract deep features on local patches and a Transformer block to learn global-context information among local patches. 
    The backbone can be any encoder and decoder structure for segmenting on local patches and we have evaluated three backbones of U-Net~\cite{ronneberger2015u}, Attention U-Net~\cite{schlemper2019attention} and U-Net++~\cite{zhou2019unet}, yielding three variations of U-Netmer.
    \item U-Netmer is a flexible model which can segment the input image with different patch sizes with identical structure and the same parameters. Jointly training the U-Netmer with different patch sizes can solve the scale-sensitivity problem. Such crafted design and astutely devised training strategies of U-Netmer allow the network to seamlessly imbibe and incorporate bountiful multi-scale contextual knowledge in learning procedures. Therefore, U-Netmer consistently provide better results on 7 public datasets compared to baselines and state-of-the-art models for segmentation.
    \item U-Netmer can also output the segmentation maps with different scales and the discrepancy of these outputs is linearly correlated to the segmentation accuracy, which can be considered as a confidence score indicating the confidence of the segmentation map and ranks the test images by the difficulty.
\end{itemize}

%2, Local information on local, global on global
%3, flexible for different scales
%1, Efficient for parallel computing

\section{Method}
%This section presents the structure of U-Netmer and variations.

\subsection{Basic structure of U-Netmer}

\begin{figure*}[!t]
\centering
\includegraphics[width=\textwidth]{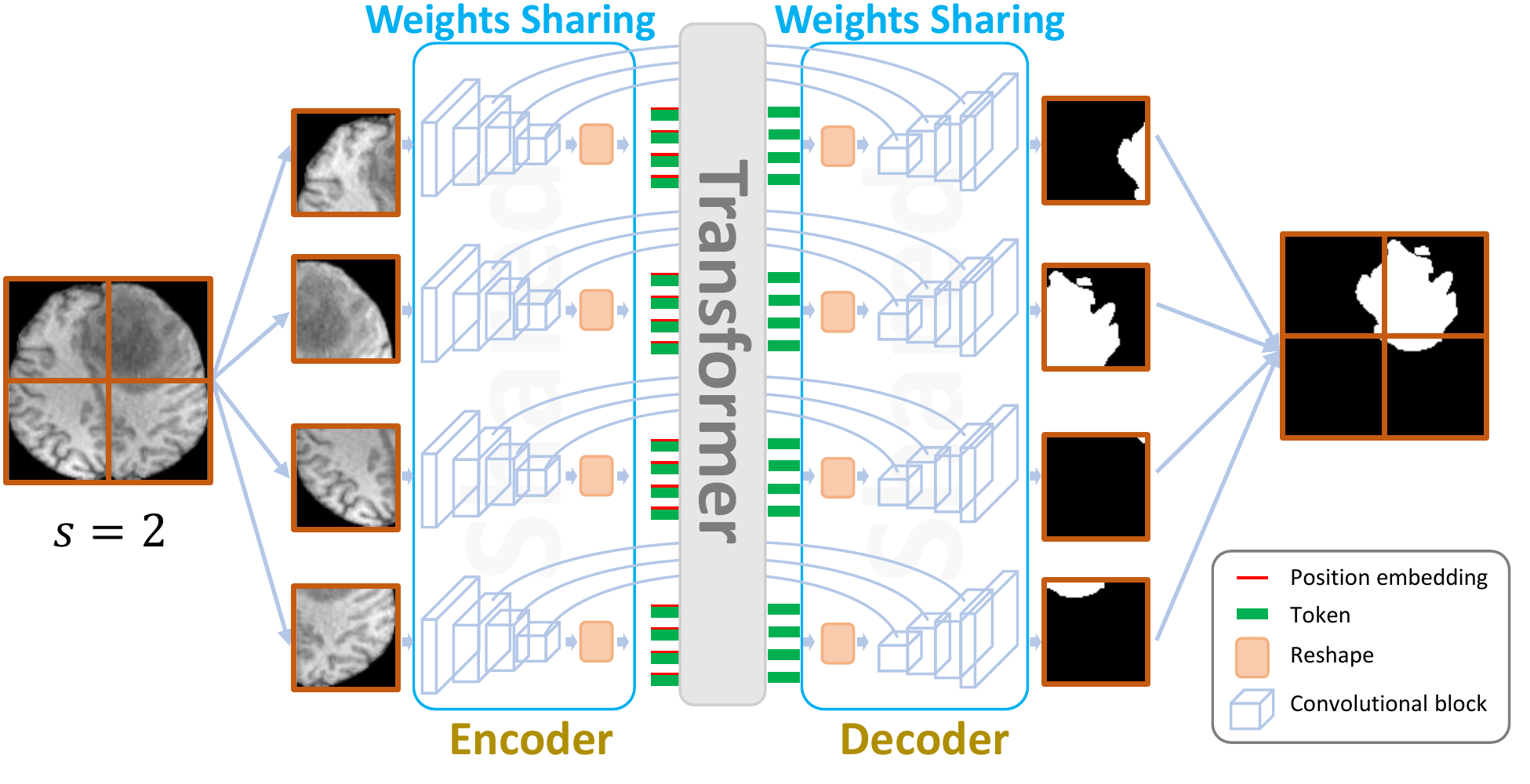}
\caption{Framework of the U-Netmer with an example of the scale $s=2$, indicating 2 patches on each side. The input image is first split into 4 local patches and each local patch is encoded into tokens by an encoder. Tokens of all local patches are fed into Transformer for learning the global context among patches.
The global-context enhanced tokens are then decoded by a decoder with the integrated information from the encoder for output prediction on each local patch.
The weights on the encoder and decoder are shared among all local patches.}
\label{fig:framework}
\end{figure*}

Fig.~\ref{fig:framework} shows the framework of the U-Netmer (with single scale $s=2$ as an example).
U-Netmer can be denoted as $\mathcal{M}_s=(\mathcal{P}_s,\mathcal{E},\mathcal{T},\mathcal{D})$, which consists of ``patchification" $\mathcal{P}_s$, encoder $\mathcal{E}$, Transformer $\mathcal{T}$,  and decoder $\mathcal{D}$, where $s\in[1,2,4,8]$ is the scale (see Fig.~\ref{fig:concepts}(c)).
Given the input image $x$ with the size of $h\times w$ ($h$ is height and $w$ is width, 2D image as an example), the segmentation output $y$ can be computed by: $y=\mathcal{M}_s(x)=\mathcal{D}(\mathcal{T}(\mathcal{E}(\mathcal{P}_s(x))))$.
Each operation is described in the following sections.

\subsubsection{Patchification $\mathcal{P}_s$}

Patchification cuts the input image into (typically equal-sized and non-overlapping) patches which is an important step in vision Transformer~\cite{dosovitskiy2020image}.
Let $s$ be the number of the patches on one side of the input image and the output of the $\mathcal{P}_s$ is a set of $s\times s$ patches (for 2D input image as an example): $\textbf{p}=\mathcal{P}_s(x)$.
The size of each local patch is $h/s \times w/s$.
Fig.~\ref{fig:concepts}(c) shows the examples of patchification with different scales $s=1,2,4,8$.

\subsubsection{Encoder $\mathcal{E}$} 

In vision Transformer~\cite{dosovitskiy2020image}, the $i$th patch $p_i\in\textbf{p}, i=1,2,..., s\times s$ is flatten into 1D feature vector.
However, for medical image segmentation, the aim is to make predictions on each pixel within local patches.
Thus, we use a segmentation backbone to convert the patches into deep features instead of directly converting the patches into 1D tokens.
The aim of using segmentation backbone is to learn the rich information among pixels within local patches for the pixel-level prediction.
Given the local patch $p_i\in\textbf{p}$, the encoder outputs $\textbf{f}_i,\tau_i=\mathcal{E}(p_i)$, where $\textbf{f}$ is a set of intermediate deep features while $\tau_i$ is the  deep feature from the last layer which contains the deep abstract and contextual information of the input local patch.
Any encoder block can be applied here to extract the deep features $\textbf{f}$ and $\tau$, such as the encoder part of the U-Net~\cite{ronneberger2015u} and U-Net++~\cite{zhou2019unet}, 
which usually consists of several convolutional layers, followed by Rectified Linear Unit (ReLU), Batch Normalization and Max-pooling layers.
The size of $\tau_i$ is $h/(2^ns)\times w/(2^ns)$ where $n$ is the number of max-pooling layer in the encoder $\mathcal{E}$.
As shown in Fig.~\ref{fig:framework}, the encoder is shared for all local patches, which can be efficiently computed in parallel.

\subsubsection{Transformer $\mathcal{T}$}

Although it is efficient to segment small local patches $\textbf{p}$, the global-context information of the input image is missed when splitting the image into local patches.
To solve this problem, Transformer~\cite{vaswani2017attention} is used to learn the global-context information among the local patches to enhance the segmentation on each local patch and further improve the accuracy of the final segmentation.
We first reshape the deep feature $\tau_i$ obtained from the encoder into a sequence of $h/(2^ns)\times w/(2^ns)$.
Note that the number of tokens does not vary with the scale $s$.
The tokens from all local patches are concatenated as 1D sequences $\bm{\tau}=[\tau_1,\tau_2,...,\tau_{s\times s}]$ with the number of $s\times s\times h/(2^ns)\times w/(2^ns)=(hw)/2^n$ tokens, indicating the number of tokens does not related to the scale $s$.
To keep the position information of local patches, a learnable position embedding vector $\nu$ is added to tokens: $\bm{\tau}=\bm{\tau}+\nu$, which is fed into the standard Transformer block~\cite{vaswani2017attention} with a multi-head self-attention (MSA) $\hat{\bm{\tau}}_l=\text{MSA}(\hat{\bm{\tau}}_{l-1})+\bm{\tau}_{l-1}$ and a feed-forward network (MLP):  
$\hat{\bm{\tau}}_l=\text{MLP}(\hat{\bm{\tau}}_l)+\hat{\bm{\tau}}_l$ where $l$ is the number of Transformer block and $\hat{\bm{\tau}}_0=\bm{\tau}$.
The detailed information on the multi-head self-attention (MSA) and feed-forward networks (MLP) can be found in studies~\cite{vaswani2017attention,dosovitskiy2020image}.
The output of Transformer $\hat{\bm{\tau}}=\mathcal{T}(\bm{\tau})$ contains the global-context information among all local patches learned by the self-attention mechanism.

\subsubsection{Decoder $\mathcal{D}$}
Similar to the encoder $\mathcal{E}$, the decoder aims to fuse the intermediate feature $f_i$ and global-context embedded feature $\hat{\tau}_i\in \hat{\bm{\tau}}$ to segment each pixel on the local patch $p_i$.
The output of the decoder $\mathcal{D}$ is the segmentation result $o_i=\mathcal{D}(\hat{\tau}_i,f_i)$.
All the outputs of local patches are stitched together as the final segmentation map $B$ of the input image $x$.
The detailed structure of the decoder $\mathcal{D}$ is related to the encoder $\mathcal{E}$, which also consists several convolutional layers, followed by Rectified Linear Unit (ReLU), Batch Normalization and Up-pooling layers.
Any decoder block of the typical segmentation neural networks can be applied, such as the encoder part of the U-Net~\cite{ronneberger2015u}, attention U-Net~\cite{schlemper2019attention}, and U-Net++~\cite{zhou2019unet}.
As shown in Fig.~\ref{fig:framework}, the decoder is also shared among all local patches which can be efficiently computed in parallel.

\subsection{Variations of U-Netmer}

As discussed before, the encoder $\mathcal{E}$ and decoder $\mathcal{D}$ can be obtained from any segmentation backbone.
In this paper, we use encoders and decoders from three typical segmentation neural networks: U-Net~\cite{ronneberger2015u}, attention U-Net~\cite{schlemper2019attention}, and U-Net++~\cite{zhou2019unet}, yielding to the U-Netmer, attention U-Netmer, and U-Netmer++, respectively.
Any other Transformer can be used as Transformer $\mathcal{T}$ block to learn the global-context information among local patches.
In this paper, we use the standard Transformer block for simplification and generalization.

\subsection{Joint training U-Netmer with multiple scales}

\begin{figure}[!t]
\centering
\includegraphics[width=0.5\textwidth]{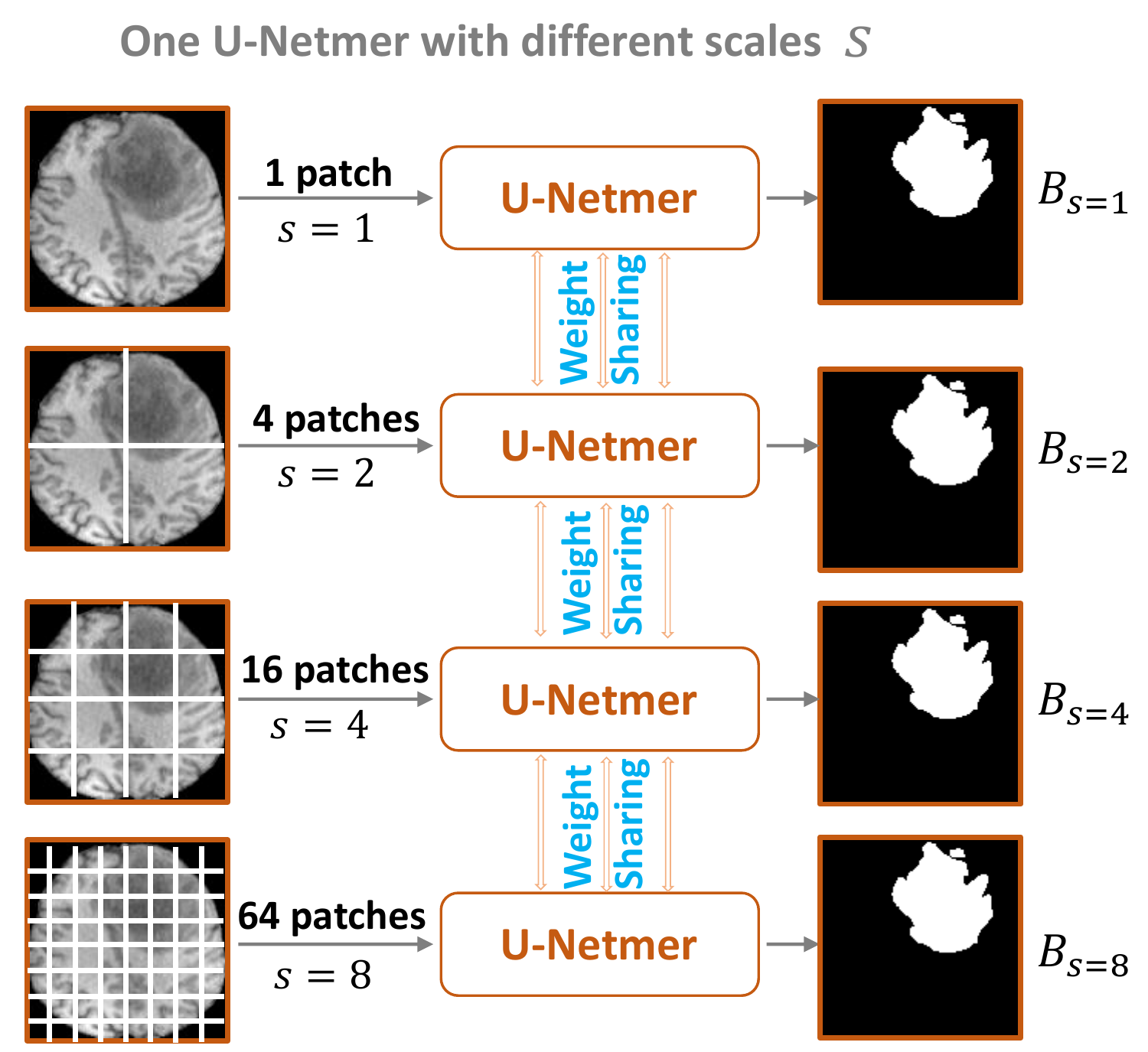}
\caption{An illustration of U-Netmer at different scales. The identical U-Netmer (with the same structure and parameters) which can be trained and tested with different scales $s$. $B_{s=i}$ is the segmentation map on scale $s=i$ (where $i\in[1,2,4,8]$).}
\label{fig:multiscale}
\end{figure}

Segmentation accuracy is sensitive to patch scale $s$ (as show in Fig.~\ref{fig:concepts}(d)).
To overcome this limitation, we train the U-Netmer with different scale $s$.
The reasons are that (1) 
the encoder $\mathcal{E}$ and decoder $\mathcal{D}$ can compute the deep features on any size of local patches (with a minimal size of $2^n$ due to the $n$ number of max-pooling layers) and (2) the number of tokens $(hw)/2^n$ in Transformer $\mathcal{T}$ does not rely on the scale $s$, indicating that cutting the input image into local patches with different scales $s$ results in the same number of tokens.
As shown in Fig.~\ref{fig:multiscale}, the same model can be trained with different scales $s=1,2,4,8$ with no added change and cost. 
Thus, the U-Netmer can learn the information across different patch sizes to boost the segmentation accuracy with an identical setup.
The patch size ($h/s\times w/s$) is small when the scale size is large and we only consider the scale values $s=1,2,4,8$ for easy computation.

In the following sections, we use ${s=\langle i|j|...\rangle}$ to denote the U-Netmer which is trained on all local patches segmented with multiple scales $i$, $j$ and others where $i<j$ and $i,j\in [1,2,4,8]$.
For example, U-Netmer++$_{s=\langle 1|2|4\rangle}$ is the U-Netmer++ trained with all local patches split with scales of $s=1,2,4$ from the input image.
U-Netmer$_{s=\langle 2\rangle}$ means the U-Netmer is only trained on local patches split with single scale $s=\langle 2\rangle$.

\section{Experiments}

\subsection{Datasets}
\label{sec:datasets}

\begin{figure}[!t]
\centering
\includegraphics[width=0.5\textwidth]{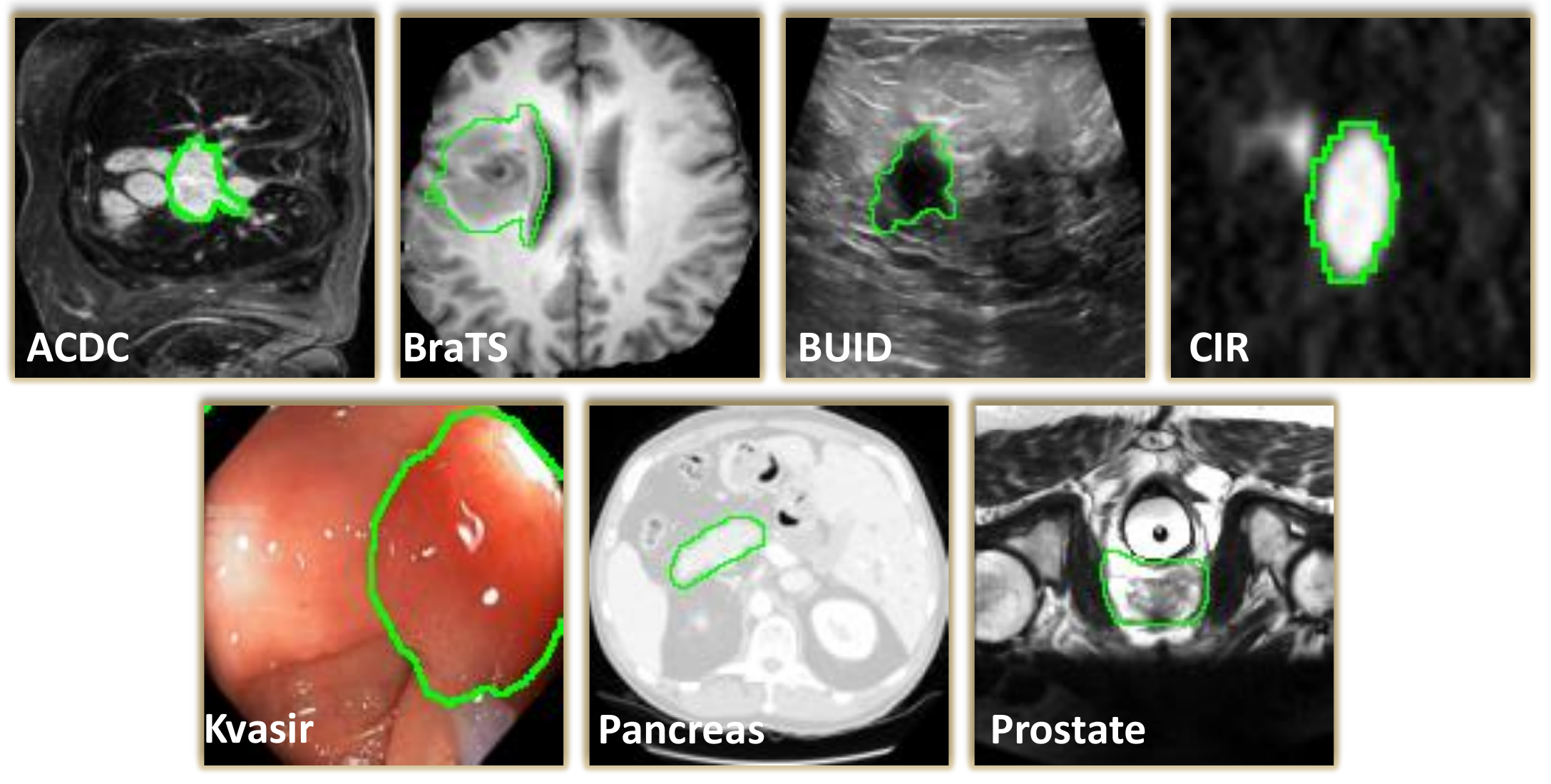}
\caption{Examples of images in the 7 datasets used in the experiments. The green contours are the boundary of ground-truth. The 7 public datasets contain images from 7 organs (brain, heart, breast, lung, poly, pancreas, and prostate) and 4 imaging modalities (MRI, CT, Ultrasound, and endoscopy). }
\label{fig:imgexamples}
\end{figure}

To evaluate the accuracy of U-Netmer, we conduct experiments on 7 publicly available datasets for medical image segmentation.
The datasets used in the experiments include
(1) \textbf{ACDC} is from the Automated Cardiac Diagnosis Challenge~\cite{bernard2018deep} with the purpose of cardiac MRI (CMR) assessment. It consists of 150 cardiac magnetic resonance images with 100 for training and the rest of 50 for testing.
(2) \textbf{BraTS} is from 2020 Multimodal Brain Tumor Segmentation Challenge~\cite{menze2014multimodal,bakas2017advancing,bakas2018identifying} with the  purpose of segmenting brain tumor (including the peritumoral edema and tumor core) segmentation. 369 scans with four modalities (T1, T1GT, T2, FLAIR) have been split into 295 ($\approx$80\%) for training and 74 ($\approx$20\%) for testing.
(3) \textbf{BUID} is from Ultrasound \& Breast Ultrasound Images Dataset~\cite{al2020dataset} with the purpose of breast cancer segmentation. 
There are 780 images which are randomly split into training (80\%) and testing (20\%) samples.
(4) \textbf{CIR}~\cite{choi2022cirdataset} consists of 956 CT images on segmented lung nodules from two public datasets which are randomly split into training (80\%) and testing (20\%) samples.
(5) \textbf{Kvasir} is from Kvasir-Seg~\cite{jha2020kvasir} which consists of 1000 polyp images (800 for training and 200 for testing).
(6) \textbf{Pancreas}~\cite{attiyeh2018survival,antonelli2022medical} consists of 285 CT scans with the purpose of pancreatic parenchyma and mass segmentation. The dataset is randomly split into 228 (80\%) scans for training and 57 (20\%) scans for testing.
(7) \textbf{Prostate} is from a Multi-site Dataset~\cite{liu2020ms} which consists 116 prostate T2-weighted MRI from three different sites. 
The dataset is separated into 80\% and 20\% for training and testing.
For 3D images, we extract 2D slices for training which are stitched into 3D for evaluation.
For CT scans,  the intensity values are truncated to the range of 5\% and 95\% percentile to remove the irrelevant details~\cite{antonelli2022medical}.
All images are normalized with zero mean and one standard deviation.
Fig.~\ref{fig:imgexamples} shows examples of images for the 7 datasets.

\subsection{Neural network training}

All models are trained with the PyTorch package, with the Adam optimizer of an initial learning rate 0.0001 which is decayed to half after every 20 epochs.
We totally trained 100 epochs with a batch size of 16.
The cross-entropy is used as the loss function in the training and Dice score is used as the evaluation metric in the testing.
To evaluate the performance of the model itself, no data augmentation or post-processing is applied for all models.
All segmentation models, including U-Netmer and other state-of-the-art models, are trained with the same dataset and same training configuration for a fair comparison.

%FLOPs for different models?

\section{Results}
In this section, we present the ablation studies of the U-Netmer with the comparison to state-of-the-art models and its potential application for ranking the test images by difficulty without ground-truth.

\subsection{Accuracy of U-Netmer}

\subsubsection{Transformer supplements U-Net}
\label{sec:pertange}
To evaluate the importance of Transformer $\mathcal{T}$ on U-Netmer with different encoders and decoders, we train models of U-Netmer$_{s=\langle i\rangle}$, attention U-Netmer$_{s=\langle i\rangle}$ and U-Netmer++$_{s=\langle i\rangle}$ with and without Transformer on local patches segmented from input image with a single scale $s=i$.
Without Transformer $\mathcal{T}$, the U-Netmer$_{s=\langle i\rangle}$, attention U-Netmer$_{s=\langle i\rangle}$ and U-Netmer++$_{s=\langle i\rangle}$ are similar to original U-Net, attention U-Net, and U-Net++ which are applied on local patches segmented from the input image with scale $s=i$.

\begin{figure*}[!t]
\centering
\includegraphics[width=\textwidth]{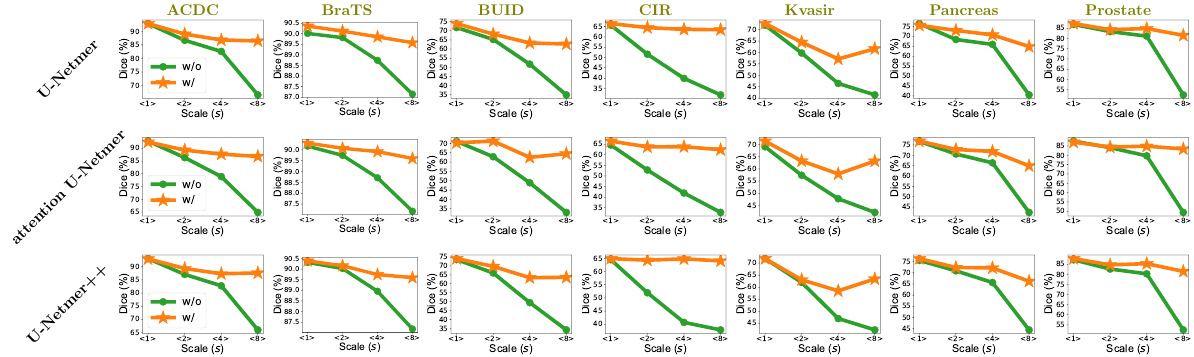}
\caption{Dice accuracy comparison of different variations of U-Netmer trained with a single scale $s=\langle i\rangle$ with Transformer $\mathcal{T}$ (w/, orange lines) and without Tranformer (w/o, green lines) on 7 datasets. The accuracy significantly drops for models without Transformer when the scale $s=i$ increases, indicating that the global-context information learned by Transformer is important for segmentation,especially with a high scale $s=\langle i\rangle$.}
\label{fig:localunet}
\end{figure*}

Fig.~\ref{fig:localunet} shows the accuracy of these models on 7 datasets.
Several observations can be obtained:
(1) Models trained with Transformer $\mathcal{T}$ have a higher accuracy than models trained without Transformer $\mathcal{T}$, especially on local patches segmented from scale $i=2,4,8$.
The results show that Transformer $\mathcal{T}$ can learn the global-context information among these patches.
The results are consistent of three different backbones (U-Net, attention U-Net and U-Net++) over 7 datasets.
(2) Unlike the visioan Transformer~\cite{dosovitskiy2020image}, splitting the input image into local patches with a single scale does not improve the accuracy for medical image segmentation on the 7 datasets and the accuracy decreases when patch sizes decrease (the scale $s$ increases) for all models with and without Transformer $\mathcal{T}$.
We also plot the percentage of the best performance of U-Netmer with Transformer $\mathcal{T}$ among different scales $s$ on test images (shown in Fig.~\ref{fig:concepts}(d)).
Results show that most test images achieve the highest accuracy on $s=1$.
For example, 85.0\%, 36.5\%, 55.4\%, 45.4\%, 54.5\%, 66.7\%, and 75.0\% of test samples achieve the best accuracy with scale $s=1$ on ACDC, BraTS, BUID, CIR, Kvasir, Pancreas, and Prostate datasets, respectively, which are larger than other scales $s=2,4,8$.
Thus, the average accuracy decreases when the scale $s$ increases in single-scale split of input images.
% D:\PaperWorks\SegmentationWork\Unetmer\results\exp01_figure_localUnet\latexfig

\subsubsection{Joint training U-Netmer with multi-scales improves the accuracy compared with single-scale split}
\begin{figure*}
\centering
\includegraphics[width=\textwidth]{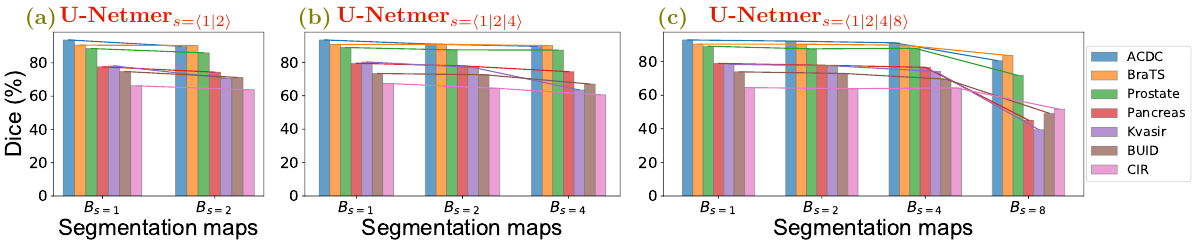}
\caption{The segmentation accuracy of different outputs $B_{s=i}$ from the joint training of (a) U-Netmer$_{s=\langle1|2\rangle}$, (b) U-Netmer$_{s=\langle1|2|4\rangle}$, and (c) U-Netmer$_{s=\langle1|2|4|8\rangle}$. }
\label{fig:segscale}
\end{figure*}

This section presents the results of the U-Netmer$_{s=\langle i|j|...\rangle}$ trained with local patches segmented with multi-scales.
If the U-Netmer is trained only on the single scale $s=\langle1\rangle$, the structure of the U-Netmer is the same to the Trans U-Net~\cite{chen2021transunet} with U-Net as the backbone.
Thus, U-Netmer$_{s=1}$ is the one baseline for comparison.
The advantage of the U-Netmer is that it can be also jointly trained with local patches segmented from multi-scales.
For example, U-Netmer$_{s=\langle1|2\rangle}$ indicates that the U-Netmer is trained on all local patches segmented with both scales $s=1$ and scale $s=2$.
We train the three variations of U-Netmer$_{s=\langle1\rangle}$ (baseline), U-Netmer$_{s=\langle1|2\rangle}$, U-Netmer$_{s=\langle1|2|4\rangle}$, and U-Netmer$_{s=\langle1|2|4|8\rangle}$.
When applying the trained U-Netmer with multi-scale patches, different segmentation outputs $B_{s=i}$ can also be obtained on the corresponding scale $s=i$ (see Fig.~\ref{fig:multiscale}).
We evaluate the accuracy of each output $B_{s=i}$ and
the results are shown in Fig.~\ref{fig:segscale}.
During testing, the accuracy of the output $B_{s=i}$ slightly decreases when scale $s=i$ increases.
For three models of U-Netmer$_{s=\langle1|2\rangle}$, U-Netmer$_{s=\langle1|2|4\rangle}$ and U-Netmer$_{s=\langle1|2|4|8\rangle}$,
the best results is achieved by the output of $B_{s=1}$ of U-Netmer jointly trained with multi-scale split, which are reported in the following sections.

%D:\PaperWorks\SegmentationWork\Unetmer\results\exp01_figure_scaleUnet\latexfig

\begin{figure*}[!t]
\centering
\includegraphics[width=\textwidth]{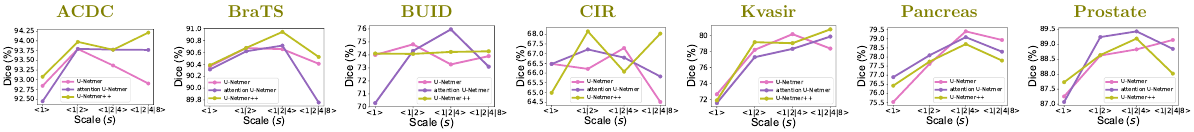}
\caption{Accuracy of U-Netmer trained with multiple scales on 7 datasets. $s=\langle1|2\rangle$ indicates that models trained on local patches segmented with scales $s=1,2$. The same definition to $s=\langle1\rangle$ (baseline), $s=\langle1|2|4\rangle$ and $s=\langle1|2|4|8\rangle$.}
\label{fig:flexibleunet}
\end{figure*}

Fig.~\ref{fig:flexibleunet} shows the accuracy of $B_{s=1}$ obtained from U-Netmer models trained with different number of scales: U-Netmer$_{s=\langle1\rangle}$, U-Netmer$_{s=\langle1|2\rangle}$, U-Netmer$_{s=\langle1|2|4\rangle}$ and U-Netmer$_{s=\langle1|2|4|8\rangle}$.
 Three variations of the U-Netmer have consistent accuracy on datasets ACDC, BraTS, Pancreas and Prostate.
For ACDC, the best accuracy is achieved by U-Netmer$_{s=\langle1|2\rangle}$ and for BraTS, Pancreas and Prostate, the best accuracy is obtained on U-Netmer$_{s=\langle1|2|4\rangle}$.
For Kvasir, U-Netmer$_{s=\langle1|2|4\rangle}$ provides the best performance with U-Net as the backbone while the highest accuracies are achieved by attention U-Netmer$_{s=\langle1|2|4|8\rangle}$ and U-Netmer++$_{s=\langle1|2|4|8\rangle}$ with attention U-Net and U-Net++ as the backbone, respectively.
A similar trend is found on BUID and CIR where the best accuracy is achieved on different scales for different backbones.
In general, the results on Fig.~\ref{fig:flexibleunet} show that training U-Netmer with multi-scales with different backbones can improve the performance on all 7 datasets, providing higher accuracies than training the U-Netmer$_{s=\langle1\rangle}$ which is the baseline model.
Table~\ref{tab:smscale} shows the accuracy of the U-Netmer with single-scale split and multi-scale split.
The results show that training the U-Netmer with multi-scale split improves the accuracy. 
Similar results have also found on attention U-Netmer and U-Netmer++.

%D:\PaperWorks\SegmentationWork\Unetmer\results\exp01_figure_scaleAccuracyUnet\latexfig

%D:\PaperWorks\SegmentationWork\Unetmer\results\exp01_figure_scaleAccuracyUnet\getTable.py
\begin{table*}[!t]
    \centering
    \caption{The Dice performance of U-Netmer with single scale $s=\langle i\rangle,i=1,2,4,8$ and multi-scale $s=\langle1|2\rangle$, $s=\langle 1|2|4\rangle$, and $s=\langle 1|2|4|8\rangle$ on 7 datasets.}
    \label{tab:smscale}
    \begin{tabular}{cl|lllllll}
    \toprule
     \multicolumn{2}{c}{Scale}  & ACDC & BUID & BraTS & CIR  & Kvasir & Pancreas & Prostate \\
     \midrule
     \multirow{4}{*}{Single-Scale}& $s=\langle1\rangle$ & 92.84{\tiny$\pm$4.11}& 74.00{\tiny$\pm$26.47}& 90.36{\tiny$\pm$5.73}& 66.48{\tiny$\pm$23.00}& 72.63{\tiny$\pm$26.56}& 75.52{\tiny$\pm$9.18}& 87.26{\tiny$\pm$3.97}\\
     & $s=\langle2\rangle$ & 88.98{\tiny$\pm$7.69}& 68.22{\tiny$\pm$28.59}& 90.13{\tiny$\pm$5.54}& 64.51{\tiny$\pm$22.08}& 64.49{\tiny$\pm$25.85}& 72.92{\tiny$\pm$10.09}& 84.29{\tiny$\pm$5.68}\\
     & $s=\langle4\rangle$ & 86.81{\tiny$\pm$11.64}& 63.33{\tiny$\pm$27.89}& 89.85{\tiny$\pm$5.88}& 63.66{\tiny$\pm$21.71}& 57.14{\tiny$\pm$27.05}& 70.69{\tiny$\pm$10.98}& 84.82{\tiny$\pm$5.26}\\
     & $s=\langle8\rangle$ & 86.46{\tiny$\pm$11.52}& 62.78{\tiny$\pm$28.89}& 89.58{\tiny$\pm$6.61}& 63.52{\tiny$\pm$21.06}& 61.63{\tiny$\pm$25.85}& 64.84{\tiny$\pm$13.10}& 81.52{\tiny$\pm$5.28}\\
     \midrule
     \multirow{3}{*}{Multi-Scale} & $s=\langle1|2\rangle$   & \textbf{93.79}{\tiny$\pm$3.30} & \textbf{74.82}{\tiny$\pm$26.40} & \textbf{90.67}{\tiny$\pm$5.25} & 66.22{\tiny$\pm$21.62} & 78.20{\tiny$\pm$22.58} & 77.62{\tiny$\pm$8.67} & 88.62{\tiny$\pm$3.52}\\
     & $s=\langle1|2|4\rangle$  & 93.37{\tiny$\pm$4.83} & 73.27{\tiny$\pm$28.61} & 90.66{\tiny$\pm$5.48} & \textbf{67.29}{\tiny$\pm$21.65} & \textbf{80.16}{\tiny$\pm$20.94} & \textbf{79.42}{\tiny$\pm$7.59} & 88.83{\tiny$\pm$3.30}\\
     & $s=\langle1|2|4|8\rangle$ 
 & 92.90{\tiny$\pm$5.04} & 73.91{\tiny$\pm$25.71} & 90.41{\tiny$\pm$5.27} & 64.51{\tiny$\pm$23.14} & 78.36{\tiny$\pm$21.22} & 78.94{\tiny$\pm$7.90} & \textbf{89.14}{\tiny$\pm$3.46}\\
     \bottomrule
    \end{tabular}
\end{table*}

%D:\PaperWorks\SegmentationWork\Unetmer\results\exp01_table_res\getTableUnetAllMetrics.py
\begin{table}[!t]
\centering
\caption{Accuracy comparison in terms of Jaccard Index, Dice, Pixel-wise accuracy, sensitivity and specificity between the U-Netmer and corresponding baselines on the 8 datasets.}
\label{tab:soabaselines}
\resizebox{0.5\textwidth}{!}{
\begin{tabular}{l|lllll}
\toprule
 ACDC & Jaccard index & Dice & Accuracy & Sensitivity & Specificity \\
\midrule
U-Net~\cite{ronneberger2015u}& 86.88{\tiny$\pm$6.42} & 92.84{\tiny$\pm$4.06} & 99.38{\tiny$\pm$0.31} & 92.04{\tiny$\pm$6.54} & 99.74{\tiny$\pm$0.16} \\
Attention U-Net~\cite{schlemper2019attention}& 86.78{\tiny$\pm$5.88} & 92.81{\tiny$\pm$3.58} & 99.37{\tiny$\pm$0.31} & 92.15{\tiny$\pm$5.89} & 99.72{\tiny$\pm$0.16} \\
U-Net++~\cite{zhou2019unet}& 87.41{\tiny$\pm$5.96} & 93.16{\tiny$\pm$3.68} & 99.41{\tiny$\pm$0.27} & 92.70{\tiny$\pm$5.69} & 99.73{\tiny$\pm$0.14} \\
Trans U-Net~\cite{chen2021transunet}& 86.89{\tiny$\pm$6.52} & 92.84{\tiny$\pm$4.11} & 99.38{\tiny$\pm$0.30} & 92.15{\tiny$\pm$6.65} & 99.73{\tiny$\pm$0.16} \\
\midrule
\rowcolor{LightCyan} U-Netmer& 88.48{\tiny$\pm$5.52} & 93.79{\tiny$\pm$3.30} & 99.46{\tiny$\pm$0.27} & \textbf{93.12}{\tiny$\pm$5.66} & 99.76{\tiny$\pm$0.13} \\
 \rowcolor{LightCyan} Attention U-Netmer& 88.55{\tiny$\pm$6.40} & 93.79{\tiny$\pm$4.01} & 99.46{\tiny$\pm$0.30} & 93.06{\tiny$\pm$6.44} & 99.77{\tiny$\pm$0.11} \\
 \rowcolor{LightCyan} U-Netmer++& \textbf{89.23}{\tiny$\pm$5.49} & \textbf{94.21}{\tiny$\pm$3.26} & \textbf{99.49}{\tiny$\pm$0.34} & 92.89{\tiny$\pm$4.73} & \textbf{99.80}{\tiny$\pm$0.20} \\
\midrule
\midrule
 BraTS & Jaccard index & Dice & Accuracy & Sensitivity & Specificity \\
\midrule
U-Net~\cite{ronneberger2015u}& 82.31{\tiny$\pm$8.66} & 90.01{\tiny$\pm$6.00} & 98.41{\tiny$\pm$0.79} & 88.33{\tiny$\pm$9.18} & \textbf{99.32}{\tiny$\pm$0.48} \\
Attention U-Net~\cite{schlemper2019attention}& 82.51{\tiny$\pm$8.12} & 90.17{\tiny$\pm$5.49} & 98.43{\tiny$\pm$0.72} & 88.74{\tiny$\pm$8.24} & 99.30{\tiny$\pm$0.48} \\
U-Net++~\cite{zhou2019unet}& 82.72{\tiny$\pm$7.96} & 90.31{\tiny$\pm$5.36} & 98.44{\tiny$\pm$0.73} & 89.12{\tiny$\pm$8.26} & 99.28{\tiny$\pm$0.49} \\
Trans U-Net~\cite{chen2021transunet}& 82.86{\tiny$\pm$8.37} & 90.36{\tiny$\pm$5.73} & 98.46{\tiny$\pm$0.76} & 89.09{\tiny$\pm$8.62} & 99.30{\tiny$\pm$0.47} \\
\midrule
\rowcolor{LightCyan} U-Netmer& 83.31{\tiny$\pm$7.82} & 90.67{\tiny$\pm$5.25} & 98.50{\tiny$\pm$0.75} & 89.38{\tiny$\pm$8.23} & \textbf{99.32}{\tiny$\pm$0.47} \\
 \rowcolor{LightCyan} Attention U-Netmer& 83.38{\tiny$\pm$7.82} & 90.72{\tiny$\pm$5.24} & 98.49{\tiny$\pm$0.75} & 90.13{\tiny$\pm$7.78} & 99.26{\tiny$\pm$0.48} \\
 \rowcolor{LightCyan} U-Netmer++& \textbf{83.76}{\tiny$\pm$7.72} & \textbf{90.95}{\tiny$\pm$5.17} & \textbf{98.52}{\tiny$\pm$0.73} & \textbf{90.37}{\tiny$\pm$7.57} & 99.29{\tiny$\pm$0.46} \\
\midrule
\midrule
 BUID & Jaccard index & Dice & Accuracy & Sensitivity & Specificity \\
\midrule
U-Net~\cite{ronneberger2015u}& 62.65{\tiny$\pm$29.13} & 71.81{\tiny$\pm$29.57} & 95.93{\tiny$\pm$5.07} & 72.84{\tiny$\pm$31.58} & 98.63{\tiny$\pm$1.77} \\
Attention U-Net~\cite{schlemper2019attention}& 61.70{\tiny$\pm$29.36} & 71.09{\tiny$\pm$29.20} & 95.87{\tiny$\pm$5.20} & 70.45{\tiny$\pm$31.22} & 98.88{\tiny$\pm$1.48} \\
U-Net++~\cite{zhou2019unet}& 63.73{\tiny$\pm$26.85} & 73.58{\tiny$\pm$26.51} & 96.00{\tiny$\pm$4.74} & 74.62{\tiny$\pm$28.23} & 98.61{\tiny$\pm$1.83} \\
Trans U-Net~\cite{chen2021transunet}& 64.33{\tiny$\pm$27.11} & 74.00{\tiny$\pm$26.47} & \textbf{96.08}{\tiny$\pm$4.93} & 74.16{\tiny$\pm$28.44} & 98.70{\tiny$\pm$1.68} \\
\midrule
\rowcolor{LightCyan} U-Netmer& 65.41{\tiny$\pm$27.17} & 74.82{\tiny$\pm$26.40} & 95.96{\tiny$\pm$5.16} & 76.47{\tiny$\pm$28.12} & 98.60{\tiny$\pm$1.78} \\
 \rowcolor{LightCyan} Attention U-Netmer& \textbf{66.35}{\tiny$\pm$25.81} & \textbf{75.97}{\tiny$\pm$24.95} & 95.95{\tiny$\pm$5.17} & \textbf{76.60}{\tiny$\pm$26.85} & 98.61{\tiny$\pm$1.88} \\
 \rowcolor{LightCyan} U-Netmer++& 64.64{\tiny$\pm$27.10} & 74.28{\tiny$\pm$26.24} & 95.90{\tiny$\pm$5.05} & 72.75{\tiny$\pm$28.65} & \textbf{98.90}{\tiny$\pm$1.46} \\
\midrule
\midrule
 CIR & Jaccard index & Dice & Accuracy & Sensitivity & Specificity \\
\midrule
U-Net~\cite{ronneberger2015u}& 52.80{\tiny$\pm$22.71} & 65.71{\tiny$\pm$23.24} & 97.57{\tiny$\pm$2.86} & 67.59{\tiny$\pm$25.51} & 98.98{\tiny$\pm$1.20} \\
Attention U-Net~\cite{schlemper2019attention}& 52.09{\tiny$\pm$24.09} & 64.62{\tiny$\pm$24.89} & 97.57{\tiny$\pm$2.90} & 66.38{\tiny$\pm$27.57} & 99.00{\tiny$\pm$1.21} \\
U-Net++~\cite{zhou2019unet}& 52.25{\tiny$\pm$24.53} & 64.57{\tiny$\pm$25.64} & 97.57{\tiny$\pm$2.93} & 65.58{\tiny$\pm$28.20} & 99.03{\tiny$\pm$1.24} \\
Trans U-Net~\cite{chen2021transunet}& 53.61{\tiny$\pm$22.58} & 66.48{\tiny$\pm$23.00} & \textbf{97.68}{\tiny$\pm$2.68} & 67.78{\tiny$\pm$25.63} & 99.04{\tiny$\pm$1.24} \\
\midrule
\rowcolor{LightCyan} U-Netmer& 54.21{\tiny$\pm$21.86} & 67.29{\tiny$\pm$21.65} & 97.60{\tiny$\pm$2.93} & \textbf{70.67}{\tiny$\pm$24.32} & 98.96{\tiny$\pm$1.25} \\
 \rowcolor{LightCyan} Attention U-Netmer& 54.13{\tiny$\pm$21.92} & 67.22{\tiny$\pm$21.61} & 97.58{\tiny$\pm$2.82} & 69.90{\tiny$\pm$24.05} & 98.99{\tiny$\pm$1.14} \\
 \rowcolor{LightCyan} U-Netmer++& \textbf{55.02}{\tiny$\pm$21.35} & \textbf{68.15}{\tiny$\pm$20.96} & 97.65{\tiny$\pm$3.02} & 70.31{\tiny$\pm$23.86} & \textbf{99.05}{\tiny$\pm$1.11} \\
\midrule
\midrule
 Kvasir & Jaccard index & Dice & Accuracy & Sensitivity & Specificity \\
\midrule
U-Net~\cite{ronneberger2015u}& 62.01{\tiny$\pm$28.11} & 71.88{\tiny$\pm$27.34} & 92.31{\tiny$\pm$10.00} & 71.91{\tiny$\pm$30.10} & 97.87{\tiny$\pm$3.69} \\
Attention U-Net~\cite{schlemper2019attention}& 59.11{\tiny$\pm$28.89} & 69.18{\tiny$\pm$28.65} & 91.92{\tiny$\pm$10.27} & 67.73{\tiny$\pm$31.57} & \textbf{98.24}{\tiny$\pm$2.87} \\
U-Net++~\cite{zhou2019unet}& 61.42{\tiny$\pm$26.88} & 71.87{\tiny$\pm$25.87} & 92.12{\tiny$\pm$9.93} & 71.62{\tiny$\pm$28.50} & 97.81{\tiny$\pm$3.38} \\
Trans U-Net~\cite{chen2021transunet}& 62.65{\tiny$\pm$27.48} & 72.63{\tiny$\pm$26.56} & 92.53{\tiny$\pm$9.80} & 71.85{\tiny$\pm$29.05} & 97.95{\tiny$\pm$3.64} \\
\midrule
\rowcolor{LightCyan} U-Netmer& 70.95{\tiny$\pm$23.64} & 80.16{\tiny$\pm$20.94} & 93.87{\tiny$\pm$8.40} & 82.80{\tiny$\pm$22.71} & 97.63{\tiny$\pm$4.26} \\
 \rowcolor{LightCyan} Attention U-Netmer& 70.57{\tiny$\pm$23.62} & 79.86{\tiny$\pm$21.17} & 93.91{\tiny$\pm$8.16} & 83.19{\tiny$\pm$23.31} & 97.42{\tiny$\pm$3.85} \\
 \rowcolor{LightCyan} U-Netmer++& \textbf{71.66}{\tiny$\pm$23.34} & \textbf{80.76}{\tiny$\pm$20.40} & \textbf{93.98}{\tiny$\pm$8.10} & \textbf{84.53}{\tiny$\pm$21.48} & 97.35{\tiny$\pm$3.96} \\
\midrule
\midrule
 Pancreas & Jaccard index & Dice & Accuracy & Sensitivity & Specificity \\
\midrule
U-Net~\cite{ronneberger2015u}& 62.51{\tiny$\pm$10.63} & 76.38{\tiny$\pm$8.53} & 99.39{\tiny$\pm$0.26} & 75.64{\tiny$\pm$12.05} & 99.74{\tiny$\pm$0.16} \\
Attention U-Net~\cite{schlemper2019attention}& 62.95{\tiny$\pm$10.08} & 76.76{\tiny$\pm$8.03} & 99.41{\tiny$\pm$0.24} & 75.18{\tiny$\pm$12.82} & 99.77{\tiny$\pm$0.13} \\
U-Net++~\cite{zhou2019unet}& 61.52{\tiny$\pm$10.56} & 75.61{\tiny$\pm$8.64} & 99.37{\tiny$\pm$0.26} & 74.88{\tiny$\pm$13.07} & 99.73{\tiny$\pm$0.16} \\
Trans U-Net~\cite{chen2021transunet}& 61.47{\tiny$\pm$10.98} & 75.52{\tiny$\pm$9.18} & 99.38{\tiny$\pm$0.25} & 73.79{\tiny$\pm$13.71} & 99.75{\tiny$\pm$0.15} \\
\midrule
\rowcolor{LightCyan} U-Netmer& \textbf{66.46}{\tiny$\pm$9.63} & \textbf{79.42}{\tiny$\pm$7.59} & \textbf{99.47}{\tiny$\pm$0.23} & \textbf{78.49}{\tiny$\pm$11.77} & 99.78{\tiny$\pm$0.11} \\
 \rowcolor{LightCyan} Attention U-Netmer& 66.02{\tiny$\pm$9.58} & 79.09{\tiny$\pm$7.63} & 99.46{\tiny$\pm$0.23} & 77.67{\tiny$\pm$11.63} & 99.79{\tiny$\pm$0.13} \\
 \rowcolor{LightCyan} U-Netmer++& 65.50{\tiny$\pm$9.69} & 78.71{\tiny$\pm$7.56} & 99.46{\tiny$\pm$0.21} & 76.62{\tiny$\pm$12.87} & \textbf{99.80}{\tiny$\pm$0.11} \\
\midrule
\midrule
 Prostate & Jaccard index & Dice & Accuracy & Sensitivity & Specificity \\
\midrule
U-Net~\cite{ronneberger2015u}& 76.91{\tiny$\pm$5.59} & 86.83{\tiny$\pm$3.74} & 98.99{\tiny$\pm$0.50} & 86.62{\tiny$\pm$6.78} & 99.58{\tiny$\pm$0.23} \\
Attention U-Net~\cite{schlemper2019attention}& 78.17{\tiny$\pm$5.57} & 87.63{\tiny$\pm$3.66} & 99.05{\tiny$\pm$0.48} & 87.64{\tiny$\pm$6.36} & 99.59{\tiny$\pm$0.21} \\
U-Net++~\cite{zhou2019unet}& 77.38{\tiny$\pm$5.30} & 87.14{\tiny$\pm$3.58} & 99.01{\tiny$\pm$0.48} & 86.60{\tiny$\pm$6.55} & 99.60{\tiny$\pm$0.22} \\
Trans U-Net~\cite{chen2021transunet}& 77.60{\tiny$\pm$5.82} & 87.26{\tiny$\pm$3.97} & 99.03{\tiny$\pm$0.47} & 86.65{\tiny$\pm$6.53} & 99.62{\tiny$\pm$0.22} \\
\midrule
\rowcolor{LightCyan} U-Netmer& 80.58{\tiny$\pm$5.39} & 89.14{\tiny$\pm$3.46} & 99.15{\tiny$\pm$0.56} & 87.01{\tiny$\pm$6.91} & \textbf{99.73}{\tiny$\pm$0.10} \\
 \rowcolor{LightCyan} Attention U-Netmer& \textbf{81.02}{\tiny$\pm$4.85} & \textbf{89.43}{\tiny$\pm$3.13} & \textbf{99.19}{\tiny$\pm$0.41} & \textbf{88.71}{\tiny$\pm$5.27} & 99.68{\tiny$\pm$0.14} \\
 \rowcolor{LightCyan} U-Netmer++& 80.63{\tiny$\pm$4.94} & 89.19{\tiny$\pm$3.13} & 99.16{\tiny$\pm$0.49} & 87.23{\tiny$\pm$6.11} & \textbf{99.73}{\tiny$\pm$0.13} \\
\bottomrule
\end{tabular}}
\end{table}

\subsubsection{U-Netmer trained with multi-scales outperforms  state-of-the-art models}

We first compare three variations of U-Netmer with their corresponding baselines:  U-Net~\cite{ronneberger2015u}, Attention U-Net~\cite{schlemper2019attention}, U-Net++~\cite{zhou2019unet} and Trans U-Net~\cite{chen2021transunet}.
Table~\ref{tab:soabaselines} shows the accuracy measured by Jaccard index, Dice coefficient, pixelwise accuracy, sensitivity and and specificity on 7 datasets.
Results show that U-Netmer outperforms its baseline models in most cases.
We also conduct the comparison study between the U-Netmer and other state-of-the-art models including other pure U-Net based models (such as BiOnet~\cite{xiang2020bio}, ConvUNeXt~\cite{han2022convunext}, ResUnet~\cite{zhang2018road}) and U-Net with Transformer models (such as UNext~\cite{valanarasu2022unext},
UCTransNet~\cite{wang2022uctransnet},
MedT~\cite{valanarasu2021medical}).
All of these models are trained with the same training setup for a fair comparison.
Table~\ref{tab:soa} shows the accuracies on the 7 datasets which shows that U-Netmer based methods (U-Netmer, Attention U-Netmer and U-Netmer++) provide higher accuracy than other models.
U-Netmer provides the highest accuracy on the Pancreas dataset, Attention U-Netmer provides the highest accuracy on the BUID, Prostate datasets and U-Netmer++ provides the highest accuracy on ACDC, BraTS, CIR, and Kvasir datasets.

\begin{table*}
\centering
\caption{The Dice scores of the U-Netmer and other state-of-the-art models on the 7 datasets.}
\label{tab:soa}
%\resizebox{\textwidth}{!}{
\begin{tabular}{l|lllllll}
\toprule
Models & ACDC & BUID & BraTS & CIR  & Kvasir & Pancreas & Prostate \\
\midrule
 U-Net~\cite{ronneberger2015u} & 92.84{\tiny$\pm$4.06} & 71.81{\tiny$\pm$29.57} & 90.01{\tiny$\pm$6.00} & 65.71{\tiny$\pm$23.24} &  71.88{\tiny$\pm$27.34} & 76.38{\tiny$\pm$8.53} & 86.83{\tiny$\pm$3.74} \\
 Attention U-Net~\cite{schlemper2019attention} & 92.81{\tiny$\pm$3.58} & 71.09{\tiny$\pm$29.20} & 90.17{\tiny$\pm$5.49} & 64.62{\tiny$\pm$24.89} & 69.18{\tiny$\pm$28.65} & 76.76{\tiny$\pm$8.03} & 87.63{\tiny$\pm$3.66} \\
 U-Net++~\cite{zhou2019unet} & 93.16{\tiny$\pm$3.68} & 73.58{\tiny$\pm$26.51} & 90.31{\tiny$\pm$5.36} & 64.57{\tiny$\pm$25.64} &  71.87{\tiny$\pm$25.87} & 75.61{\tiny$\pm$8.64} & 87.14{\tiny$\pm$3.58} \\
 BiOnet~\cite{xiang2020bio} & 91.01{\tiny$\pm$4.93} & 67.91{\tiny$\pm$32.27} & 90.51{\tiny$\pm$4.91} & 65.77{\tiny$\pm$22.72} & 68.29{\tiny$\pm$29.08} & 77.59{\tiny$\pm$8.49} & 86.39{\tiny$\pm$4.71} \\
 ConvUNeXt~\cite{han2022convunext} & 85.28{\tiny$\pm$8.59} & 64.21{\tiny$\pm$27.79} & 88.23{\tiny$\pm$6.67} & 60.97{\tiny$\pm$25.43} &  49.30{\tiny$\pm$26.47} & 56.25{\tiny$\pm$12.60} & 80.46{\tiny$\pm$6.49} \\
 ResUnet~\cite{zhang2018road} & 91.94{\tiny$\pm$6.08} & 57.91{\tiny$\pm$29.84} & 90.29{\tiny$\pm$4.93} & 58.03{\tiny$\pm$28.39} &  67.66{\tiny$\pm$23.35} & 76.31{\tiny$\pm$8.78} & 87.01{\tiny$\pm$4.42} \\
UNext~\cite{valanarasu2022unext} & 86.32{\tiny$\pm$7.63} & 64.12{\tiny$\pm$30.62} & 89.23{\tiny$\pm$5.90} & 58.31{\tiny$\pm$28.61} &  57.26{\tiny$\pm$25.50} & 56.08{\tiny$\pm$11.73} & 81.86{\tiny$\pm$5.99} \\
 UCTransNet~\cite{wang2022uctransnet} & 93.39{\tiny$\pm$4.19} & 72.39{\tiny$\pm$28.12} & 90.53{\tiny$\pm$5.77} & 64.79{\tiny$\pm$23.99} &  78.25{\tiny$\pm$24.05} & 75.31{\tiny$\pm$8.99} & 88.57{\tiny$\pm$3.31} \\
 Trans U-Net~\cite{chen2021transunet} & 92.84{\tiny$\pm$4.11} & 74.00{\tiny$\pm$26.47} & 90.36{\tiny$\pm$5.73} & 66.48{\tiny$\pm$23.00} &  72.63{\tiny$\pm$26.56} & 75.52{\tiny$\pm$9.18} & 87.26{\tiny$\pm$3.97} \\
 MedT~\cite{valanarasu2021medical} & 77.41{\tiny$\pm$12.03} & 48.67{\tiny$\pm$29.51} & 89.91{\tiny$\pm$6.29} & 59.73{\tiny$\pm$26.29} &  37.69{\tiny$\pm$29.27} & 43.27{\tiny$\pm$14.87} & 74.65{\tiny$\pm$5.93} \\
 \midrule
 \rowcolor{LightCyan} U-Netmer & 93.79{\tiny$\pm$3.30} & 74.82{\tiny$\pm$26.40} & 90.67{\tiny$\pm$5.25} & 67.29{\tiny$\pm$21.65} & 80.16{\tiny$\pm$20.94} & \textbf{79.42}{\tiny$\pm$7.59} & 89.14{\tiny$\pm$3.46} \\
 \rowcolor{LightCyan} Attention U-Netmer & 93.79{\tiny$\pm$4.01} & \textbf{75.97}{\tiny$\pm$24.95} & 90.72{\tiny$\pm$5.24} & 67.22{\tiny$\pm$21.61} &  79.86{\tiny$\pm$21.17} & 79.09{\tiny$\pm$7.63} & \textbf{89.43}{\tiny$\pm$3.13} \\
 \rowcolor{LightCyan} U-Netmer++ & \textbf{94.21}{\tiny$\pm$3.26} & 74.28{\tiny$\pm$26.24} & \textbf{90.95}{\tiny$\pm$5.17} & \textbf{68.15}{\tiny$\pm$20.96} & \textbf{80.76}{\tiny$\pm$20.40} & 78.71{\tiny$\pm$7.56} & 89.19{\tiny$\pm$3.13} \\
\bottomrule
\end{tabular}
\end{table*}

\subsubsection{Deep features learned by U-Netmer have higher segmentation ability than those from its counterpart (U-Net)}
\begin{figure*}[!t]
\centering
\includegraphics[width=\textwidth]{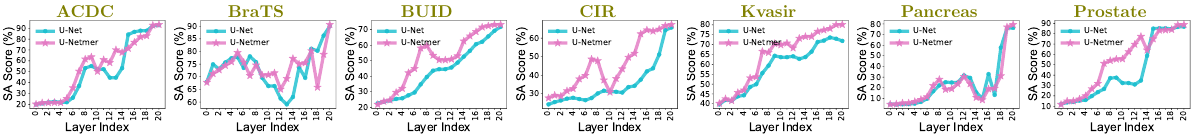}
\caption{The segmentation ability (SA) scores of the U-Net and U-Netmer$_{s=<1|2|4>}$ on different layers (there are 18 layers on U-Net indexed from 1 to 19) measured by the ProtoSeg~\cite{he2023segmentation}. The input image is indexed as 0 while the segmentation output is indexed as 20.}
\label{fig:protoseg}
\end{figure*}

This section presents the segemtnation ability (SA) scores of each layer of the U-Netmer$_{s=\langle1|2|4\rangle}$ and the original U-Net computed by the ProtoSeg~\cite{he2023segmentation} which can compute the binary segmentation map on deep features based on prototypes of background and target regions.
There are 18 layers in a typical U-Net and the U-Netmer with U-Net as the backbone.
The ProtoSeg can also be applied on input image which is considered as a specifical feature map~\cite{he2023segmentation}.
Fig.~\ref{fig:protoseg} shows the SA scores of U-Net and U-Netmer including input images (index 0), 18 deep features (index 1-19) and the final segmentation output (index 20).
It can be seen from the figure that the segmentation ability scores of the U-Netmer trained with different patch sizes are higher than the corresponding SA scores of the U-Net trained with the whole input image, especially on ACDC, BUID, CIR, Kvasir, and Prostate datasets.
The results show that U-Netmer trained with multi-scales can improve the segmentation ability on deep features as well as on the final output.

%D:\PaperWorks\SegmentationWork\Unetmer\results\exp03_protoseg_curves\latexfig

\subsubsection{Discussion}

Experimental results show that the vanilla U-Net~\cite{ronneberger2015u} provides a moderate accuracy based on small input patches which contain limited information for separating the target region and background.
Applying a Transformer block among the local patches segmented from the same input image can significantly improve the accuracy (see Fig.~\ref{fig:localunet}) because the local patches can learn the global-context information by the self-attention mechanism in Transformer~\cite{he2021global}.
This inspires the design of the U-Netmer which can make a prediction based on local patches with different patch sizes without changes to the model structure and parameters.
However, training the U-Netmer with a single scale (a fixed patch size) does not improve the accuracy (see Fig.~\ref{fig:localunet}) due to the fact that segmentation is sensitive to the scales.
To solve this problem, we train the U-Netmer with local patches segmented with multi-scales which can improve the accuracies for segmentation. The same results are also found on the three different variations (see Fig.~\ref{fig:flexibleunet}).
The U-Netmer trained by patches with different sizes (multi-scales) provides the highest accuracy compared to four different baselines and other U-Net based and U-Net combined with Transformer models (see Table~\ref{tab:soa}),
indicating that training neural networks with multiscale information can improve the accuracy of segmentation.
By computing the segmentation ability scores~\cite{he2023segmentation} on layers of the U-Netmer, we find that the training U-Netmer with different patch sizes can improve the segmentation ability on different layers and further improve the final segmentation.
Overall, experimental results have shown that U-Netmer trained with multi-scales can improve the accuracy on the test 7 datasets with different modalities.

\subsection{U-Netmer can be used to rank test images by difficulty without ground-truth}

%D:\PaperWorks\SegmentationWork\Unetmer\results\exp04_imageranking_vis\RankingDemoFig
\begin{figure}
\centering
\includegraphics[width=0.5\textwidth]{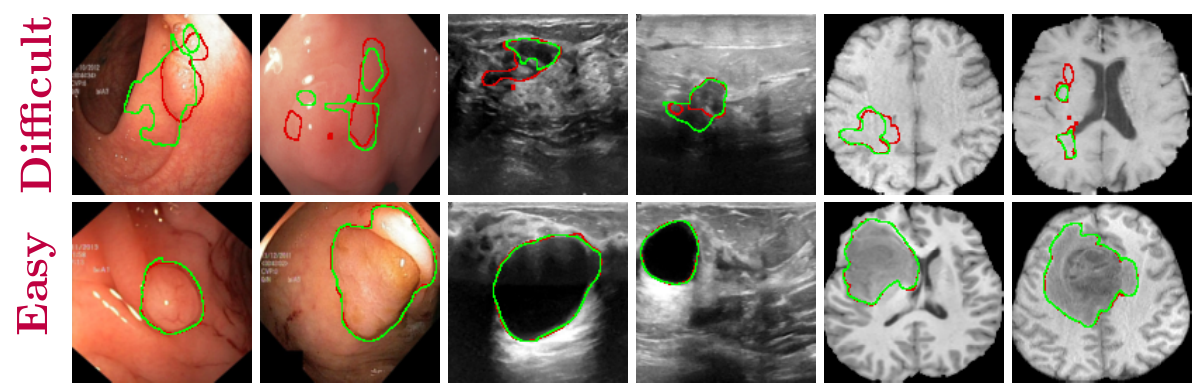}
\caption{Examples of difficult and easy examples for segmentation. The red contours are the segmentation results of U-Netmer with scale $s=1$ and the green contours are the segmentation results with scale $s=2$.}
\label{fig:exmaplesr}
\end{figure}

%D:\PaperWorks\SegmentationWork\Unetmer\results\exp04_protoseg_rankingig
\begin{figure}[!t]
\centering
\includegraphics[width=0.5\textwidth]{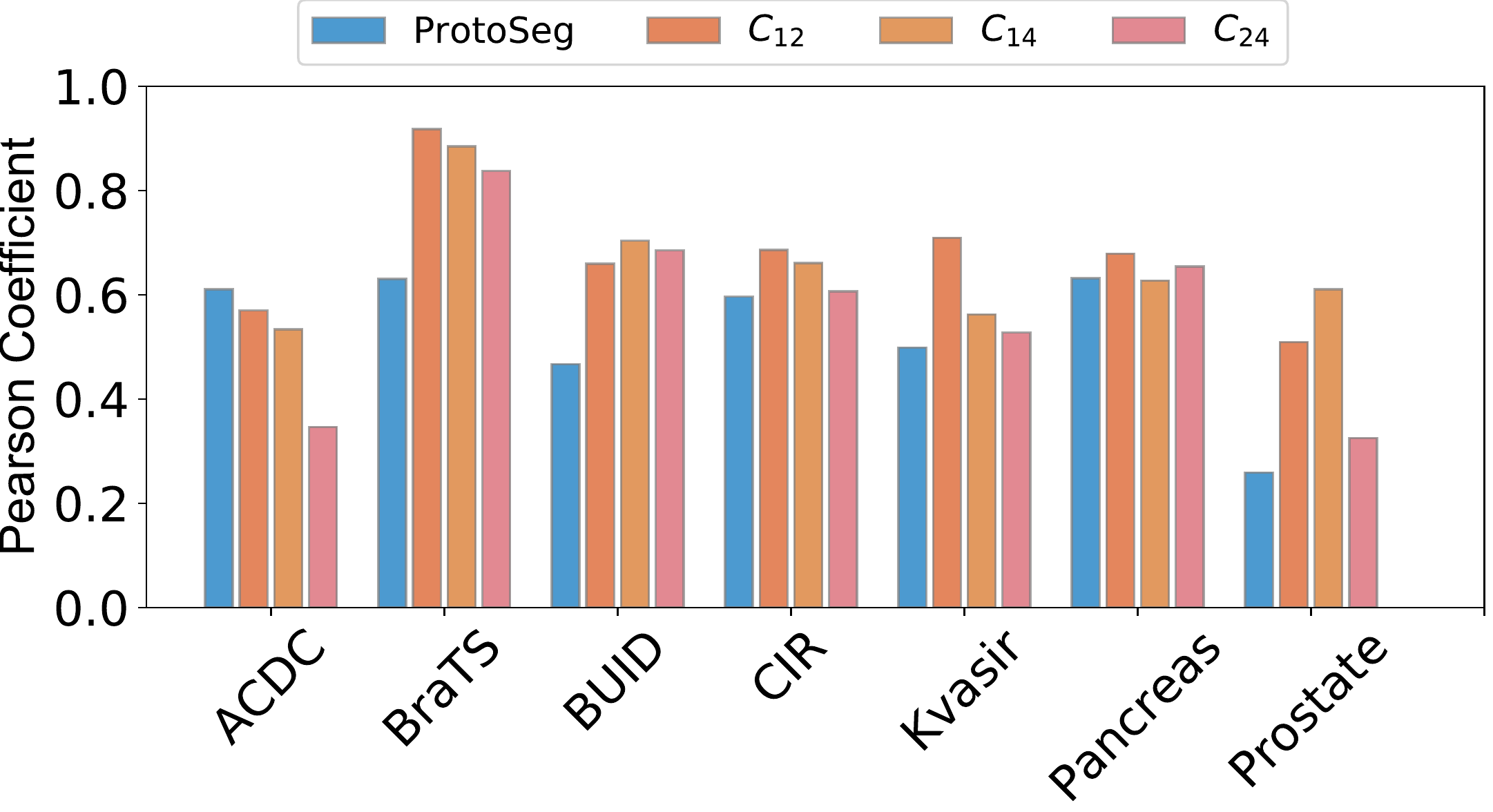}
\caption{Comparison of the Pearson coefficient between the segmentation
accuracy and confidence scores computed by ProtoSeg~\cite{he2023segmentation} and $\mathcal{C}_{12}$, $\mathcal{C}_{14}$, ad $\mathcal{C}_{24}$ from U-Netmer$_{s=\langle1|2|4\rangle}$.}
\label{fig:pearsonr}
\end{figure}

%D:\PaperWorks\SegmentationWork\Unetmer\results\exp04_protoseg_ranking\latexfig
\begin{figure*}[!ht]
\centering
\includegraphics[width=\textwidth]{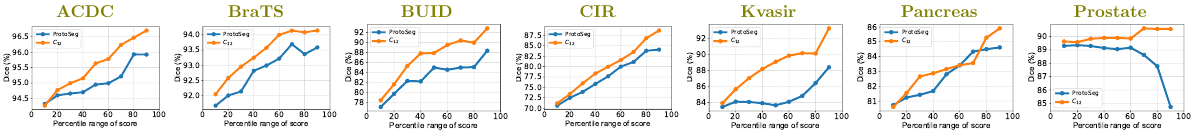}
\caption{ The segmentation accuracy (y-axis) for the test images thresholded by the confidence score (x-axis) computed by ProtoSeg~\cite{he2023segmentation} and U-Netmer.}
\label{fig:percentagerank}
\end{figure*}

%D:\PaperWorks\SegmentationWork\Unetmer\results\exp04_imageranking_vis
\begin{figure}[!ht]
\centering
\includegraphics[width=0.5\textwidth]{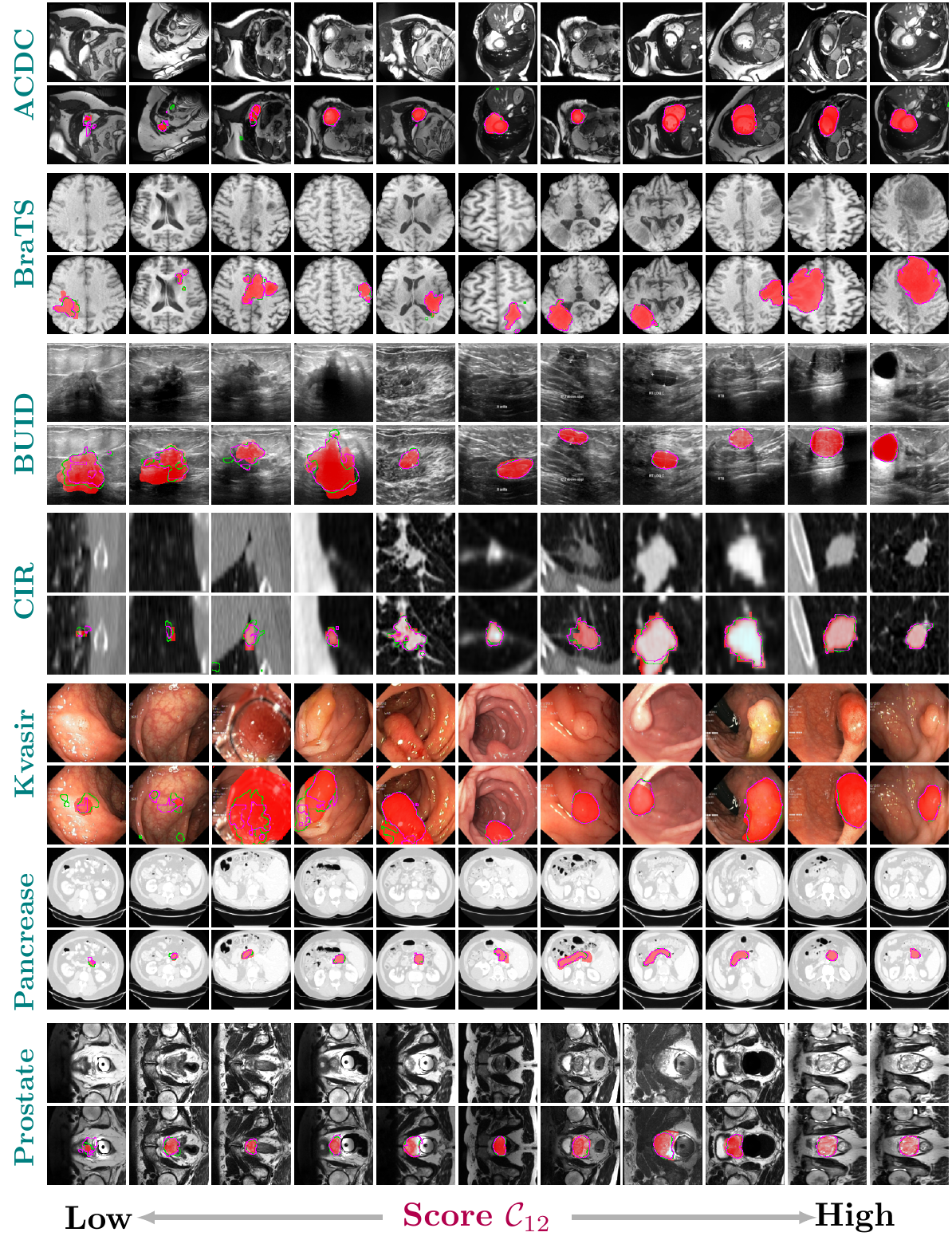}
\caption{Examples of the images (the first row on each data set) and their corresponding segmentation results (the second row on each dataset) ranked by
the confidence score $\mathcal{C}_{12}$. The red masks denote the ground-truth and the green and pink contours denote the segmentation results of the $B_{s=1}$ and $B_{s=2}$, respectively.
}
\label{fig:visulization}
\end{figure}

Ranking the test images by difficulty without ground-truth can provide useful information for end users to automatically identify the most challenging examples for human experts to review~\cite{agarwal2022estimating,he2023segmentation}.
Most segmentation models only output the segmentation result 
without such a ``confidence" evaluation.
This limits the use of segmentation algorithms in clinical practice when certain acceptance criteria are required~\cite{budd2021survey} or when we need to prioritize users’ time on inspection and auditing~\cite{agarwal2022estimating}.
There is an unmet need to evaluate segmentation
accuracies and even to reject failed segmentations in the real-world applications  when the ground-truth is absent in reality~\cite{valindria2017reverse}. 
Most studies estimate the pixel/voxel level uncertainty~\cite{agarwal2022estimating,shi2021inconsistency,wickstrom2020uncertainty} to highlight the challenging regions within
the target region or background on the single input image.
These methods do not provide the image level confidence scores to automatically select a subset of challenging samples.

The U-Netmer can provide an estimation of the confidence score which can be used to rank the test image by difficulty without ground-truth.
Given a test image, U-Netmer can make predictions with different scales.
We use $B_{s=i}$ to denote the segmentation map computed with scale $s=i$.
Fig.~\ref{fig:exmaplesr} shows testing examples of the segmentation maps from the scale $s=1$ ($B_{s=1}$) and $s=2$ ($B_{s=2}$) of the trained U-Netmer$_{s=\langle1|2|4\rangle}$.
One observation is that the segmentation maps of $B_{s=1}$ and $B_{s=2}$ are similar on easy samples while they are different on difficult samples.
Part of the reason is that the target regions (e.g., lesions) on difficult samples have low contrast information compared to their background or have a small size, yielding to different segmentation results with different scales.
Therefore, their discrepancy can be used to measure the difficulty of the test images or rank the test images by difficulty without ground-truth for segmentation.

To measure the difficulty of the test images without ground-truth, we
compute the discrepancy between the estimation of $B_{s=i}$ and $B_{s=j}$ (where $i\neq j$) by: $\mathcal{C}_{ij}=\mathcal{D}\big(B_{s=i},B_{s=j}\big)$, 
where $\mathcal{D}$ is a distance metric to measure the difference between
two estimated segmentation maps of $B_{s=i}$ and $B_{s=j}$. $\mathcal{C}$ has a
different meanings given different distance metric $\mathcal{D}$. If the
distance metric is defined as 
$\mathcal{D}=|B_{s=i}-B_{s=j}|$, the $\mathcal{C}$ indicates
the uncertainty regions of the segmentation map~\cite{nair2020exploring}. In this
paper, we use the Dice coefficient as the distance metric to
measure the difficulty to segment each image, defined as: $\mathcal{C}_{ij}=2|B_{s=i}\cap B_{s=j}|/|B_{s=i}+B_{s=j}|$
where $\mathcal{C}_{ij}$ measures the consistency between the segmentation
maps between $B_{s=i}$ and $B_{s=j}$ obtained from U-Netmer with different scales $s=i$ and $s=j$.
$\mathcal{C}_{ij}$ is considered as the confidence score to estimate the segmentation accuracy of the testing images without ground-truth for end users.

Fig.~\ref{fig:pearsonr} shows the Pearson correlation between the Dice accuracy of the final segmentation and the confidence score $\mathcal{C}_{ij}$ obtained by U-Netmer$_{s=\langle1|2|4\rangle}$.
We also compare them with the baseline of the mean SA score computed by ProtoSeg~\cite{he2023segmentation} which is the mean segmentation ability score computed on the last two layers of the neural network for ranking the test images by difficulty.
The results show that the Pearson correlation between $\mathcal{C}$ and the segmentation accuracy is higher than the one between ProtoSeg and the segmentation accuracy on 6 datasets except on ACDC dataset.
In addition, $\mathcal{C}_{12}$ provides the highest correlation on BraTS, CIR, Kvasir, Pancreas while $\mathcal{C}_{14}$ provides the highest correlation on the BUID and Prostate datasets.

To test whether the confidence score obtained by U-Netmer is discriminative between easy and difficult test
samples, we plot the segmentation accuracy of test samples
bucketed by the decile of confidence scores~\cite{agarwal2022estimating}. 
We first
rank all testing samples based on the estimated confidence score and the testing samples within the $d$\% percentile are
included to compute the accuracy of the segmentation. This
is similar to the coverage~\cite{ghesu2021quantifying} which rejects the (100-$d$)\%
difficult samples for further attention. 
Fig.~\ref{fig:percentagerank} shows the accuracy of segmentation with different percentile $d$.
We show that examples at the highest percentiles on the
rank often have high segmentation accuracy and the scores computed by U-Netmer provide higher accuracy than ProtoSeg~\cite{he2023segmentation} on 6 datasets except for Pancreas.
The results also demonstrate
that the confidence scores have a high correlation with the
segmentation accuracies on the test images.

Fig.~\ref{fig:visulization} visualizes test images and their corresponding
segmentation results ranked by the confidence scores $\mathcal{C}$ on the
7 datasets. Images with low confidence scores tend to have small target regions, smooth boundaries and poor contrast between the target and background tissues. The accuracies of
their segmentation results are usually low. Images with high
confidence scores often have large target regions and clear
boundaries, yielding more consistent segmentation maps obtained from different scales of U-Netmer. For test images without
ground-truth, the confidence score can be used by end-users with a human-in-the-loop strategy: it can suggest the segmentation results of
these images with low confidence scores to the human users
for further review.

\section{Conclusion}

In conclusion, we have presented the U-Netmer which is a combination of CNN-based neural network and Transformer for medical image segmentation.
We have studied three variations of the U-Netmer where the backbones are from three typical segmentation neural networks: U-Netmer, Attention U-Netmer, and U-Netmer++.
Experimental results on 7 datasets for medical image segmentation with different modalities have shown that the U-Netmer can provide competitive results compared to four baselines and six state-of-the-art models.
The U-Nemter can also provide segmentation maps with different scales on test images. 
The discrepancy of these segmentations is linearly correlated to the segmentation accuracy, which can be considered as a confidence score to rank the test images by difficulty when the ground-truth is absent.
 This is important in real world applications, as it can highlight most difficult cases, or least accuracy cases, for users' further inspection and edit.

\bibliographystyle{IEEEtran}
\bibliography{IEEEabrv,unetmer.bib}

\end{document}